\documentclass[10pt, aps, rmp, twocolumn, superscriptaddress, floatfix]{revtex4-2}

\usepackage{mathtools}
\usepackage[english]{babel}
\usepackage{lmodern}
\usepackage{graphicx}\graphicspath{{./}{./figures}}
\usepackage{booktabs}
\usepackage{tabularx}
\usepackage{silence}
\usepackage{siunitx}

\setcitestyle{numbers, square, comma}

\usepackage[hidelinks]{hyperref}

\makeatletter
\newcommand{\subfigref}[2][]{%
  \@ifpackageloaded{hyperref}{%
    \hyperref[#2]{\autoref*{#2}#1}%
  }{%
    Figure~\ref{#2}#1%
  }%
}
\makeatother

\providecommand{\CSL}{Complex Systems Lab, Universitat Pompeu Fabra, Dr Aiguader 88, 08003 Barcelona, Spain.}
\providecommand{\ICREA}{Institució Catalana de la Recerca i Estudis Avançats (ICREA), Pg. Lluís Companys 23, 08010 Barcelona, Spain.}
\providecommand{\IBE}{Institut de Biologia Evolutiva, CSIC-UPF, Pg. Marítim de la Barceloneta 37, 08003 Barcelona, Spain.}
\providecommand{\SFI}{Santa Fe Institute, 1399 Hyde Park Road, Santa Fe, NM 87501, United States.}

\begin{document}

\title{Cognition as least action: the \textit{Physarum} Lagrangian}

\author{Ricard Sol\'{e}}
\email[Corresponding author, ]{ricard.sole@upf.edu}
\affiliation{\CSL}
\affiliation{\ICREA}
\affiliation{\IBE}
\affiliation{\SFI}

\author{Jordi Pla-Mauri}
\affiliation{\CSL}
\affiliation{\IBE}

\begin{abstract}
The slime mould Physarum polycephalum displays adaptive transport dynamics and network formation that have inspired its use as a model of biological computation. We develop a Lagrangian formulation of Physarum's adaptive dynamics on predefined graphs, showing that steady states arise as extrema of a least-action functional balancing metabolic dissipation and transport efficiency. The organism's apparent ability to find optimal paths between nutrient sources and sinks emerges from minimizing global energy dissipation under predefined boundary conditions that specify the problem to be solved. Applied to ring, tree, and lattice geometries, the framework accurately reproduces the optimal conductance and flux configurations observed experimentally. These results show that Physarum's problem-solving on constrained topologies follows a physics-based variational principle, revealing least-action dynamics as the foundation of its adaptive organization.
\end{abstract}

\maketitle

\section{Introduction}

In contrast to artificial computational systems, living agents gather, process, and store information in ways that often depart from standard circuit designs.
These natural computation systems include standard neural systems based on neurons connected by means of synaptic couplings that allow learning and distributed, emergent computation \cite{hopfield1982neural, forrest1990emergent}, particularly within the context of standard neural networks \cite{hertz2018introduction, rojas2013neural} but also liquid brains, such as ant colonies or immune systems \cite{farmer1990rosetta, sole2019liquid, pinero2019statistical, fernandez2025foraging}, where individual units (brains or immune system cells) do not maintain stable connections.
However, there are other classes of computation grounded in morphology, where the outcome of a given information process is mapped into a shape.
In addition to ants using chemical trails to organize their search patterns in space \cite{deneubourg1983probabilistic, deneubourg1990self, sole2000pattern}, plants are also known to display high morphological plasticity that allows them to adapt to environments in flexible ways \cite{oborny2019plant, trewavas2003aspects}.

\begin{figure*}[t]
    \centering
    \includegraphics[width=0.95\textwidth]{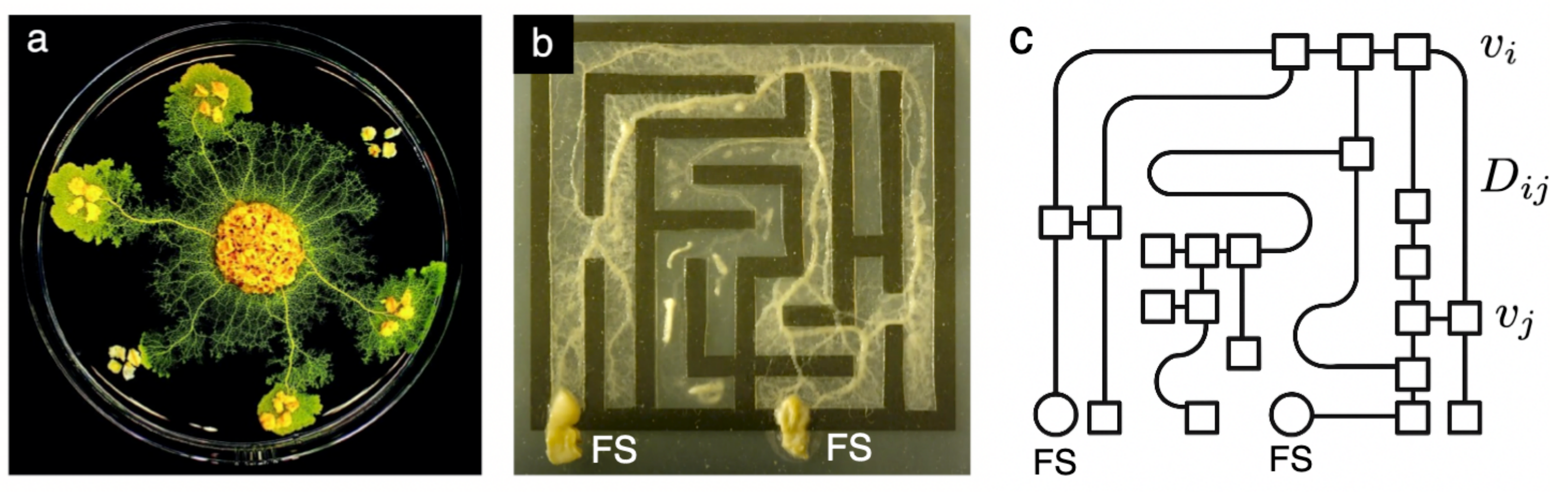}
    \caption{
        (a) {\it \textit{Physarum} polycephalum} forming a dynamic transport network.
        The plasmodium organizes its protoplasmic veins into an adaptive, spatially efficient network that optimizes nutrient distribution and environmental exploration (image courtesy of Audrey Dussutour).
        (b) Navigating a maze.
        Starting from a uniform inoculation throughout the labyrinth, the plasmodium retracts from dead ends, maintaining only the optimal path connecting food sources, thereby illustrating decentralized problem-solving through adaptive network reorganization.
        This problem has been mapped into a graph (c) and numerically solved \cite{Tero2007}, showing that the problem can be represented in terms of many coupled ordinary differential equations on a graph.
    }
    \label{Physlagr2}
\end{figure*}

A particularly revealing case study is offered by the \textit{Physarum polycephalum} slime mold, a multinucleate amoeboid organism that behaves as a single giant cell.
\textit{Physarum} belongs to a broader class of aneural agents---living systems without nervous tissues---that nevertheless display remarkable adaptive and problem-solving abilities \cite{reid2016collective,reid2023thoughts}.
These organisms are often discussed under the concept of \textit{basal cognition}, which denotes the most fundamental level of cognitive activity in life forms lacking neurons.
It encompasses basic processes such as sensing, information integration, adaptive responses, and learning-like plasticity in single cells, microbial colonies, or simple multicellular systems \citep{lyon2006biogenic, Lyon2015, BaluskaLevin2016, levin2021uncovering, levin2023darwin}.

In nature, it can reach sizes of several hundred square centimeters \cite[see also Figure 1a]{adamatzky2010physarum, adamatzky2016advances}.
Its behavior is driven by a complex adaptive network of contractile tubes.
These tubes can actively change (grow and shrink) through active mechanochemical processes as the internal medium flows through the network \cite{adamatzky2010physarum, grube2016physarum,le2024physarum}.
From a distance, the mold looks like a complex web of tubes exhibiting a constant, active remodeling in response to external cues.

The fundamental properties of this adaptive network can be captured by using a dynamical system on a graph.
This network representation has been useful in understanding the properties of a diverse range of complex systems involving transport and adaptation, including electric circuits \cite{i2001topology}, leaf venation graphs \cite{bohn2007structure, katifori2010damage, hu2013adaptation}, routing \cite{kelly1991network, guimera2002optimal, valverde2004internet} and in other areas of network science involving optimization and spatial constraints  \cite{barabasi2016network, barthelemy2011spatial}.
The response of this organism to external boundary conditions and spatially organized food sources \cite{dussutour2010amoeboid} has been exploited to solve a variety of interesting problems, including finding the shortest paths connecting exits in a maze \cite{nakagaki2000maze} (\subfigref[b--c]{Physlagr2}) as well as other mathematical problems, such as optimal transportation \cite{tero2010rules}, mapping cosmic density fields \cite{hasan2024filaments} or solving the Hanoi Tower problem \cite{reid2013solving}, the two-armed bandit problem \cite{reid2016decision} among others \cite{adamatzky2010physarum,jones2015pattern,whiting2016towards}.

How does this unicellular organism ``solve'' these problems? In nature, where no complex boundary conditions are at work, \textit{Physarum} explores the environment by extending several advancing fronts (see \subfigref[a]{Physlagr2}) that sense and respond to local chemical cues.
These fronts detect gradients of attractants released by food sources, allowing the organism to compare multiple potential routes simultaneously.
Over time, \textit{Physarum} retracts less favorable branches and reinforces the most efficient connection toward the strongest gradient.
This adaptive process effectively results in the optimization of its transport network, often described as a form of decentralized problem solving \cite{dussutour2024flow}.
This dynamics has been described within the context of fluid dynamics, molecular biophysics, and active matter and has also been shown to be displayed by other kinds of physical systems \cite{stern2023learning}.

What class of cognition encompasses the remarkable morphological complexities displayed by this giant amoeba? Is there a genuine form of ``intelligence'' underlying its ability to solve spatial problems within pre-specified boundary conditions? In this paper, we address these questions by proposing that \textit{Physarum polycephalum} operates according to a universal optimization principle long established in physics and mathematics: the \emph{principle of least action} \cite{susskind2014theoretical}.
Beyond the classical domain of physics, this principle has been used in different contexts within the biological sciences, from neurobiology \cite{senn2024neuronal}, artificial neural networks \cite{wang2022neural}, ecological systems \cite{samuelson1974biological,deng2024development}, and evolutionary dynamics \cite{kaila2008natural}.

Within this framework, the ability of plasmodium to establish efficient connections between nutrient sources without any form of centralized control or neural processing appears as a natural consequence of its material and dynamical organization.
Interestingly, this behavior is not unique to biological systems.
Comparable instances of path optimization emerge spontaneously in several non-living physical and chemical systems governed by local rules and energy minimization.
Examples include: (a) water percolating through porous or granular media, which follows paths of least hydraulic resistance \cite{Bejan1997}; (b) reaction--diffusion fronts that propagate along shortest routes in maze-like geometries \cite{steinbock1995navigating}; (c) chemotactic droplets that navigate chemical gradients to reach attractant sources \cite{lagzi2010maze}; and (d) ensembles of rolling marbles on tilted surfaces that collectively relax into minimal-energy configurations, effectively reproducing optimal transport networks \cite{adamatzky2010physarum,adamatzky2019brief}.

Although these systems lack evolved cognition, their dynamics demonstrate that simple local interactions constrained by physical laws can generate global optimization.
This suggests that the ability to discover efficient paths (whether in a living organism or an inanimate medium) emerges from fundamental variational principles rather than from computation in the conventional sense.
In this context, we propose that the least action principle, formulated through an appropriate Lagrangian representation of the \textit{Physarum} network, offers a unifying theoretical framework linking biological self-organization and physical optimization.

The paper is organized as follows.
In Section~\ref{sec:physarum_network}, we define the general problem by framing the \textit{Physarum} as a dynamical network and introduce the conceptual idea of the least action.
In Section~\ref{sec:lagrangian_steady_network}, the Lagrangian for the steady network is defined on a graph.
In Sections~\ref{sec:ring}, \ref{sec:twobranch}, and \ref{sec:lattice}, three case studies are considered, and the Lagrangian is constructed and analyzed in all three cases---namely, the ring, the branched path (both explored in \cite{Tero2007}), and the square lattice.
A general discussion, and potential extensions and analogies with other cognitive systems (such as Hebbian networks) are presented in Section~\ref{sec:discussion}.

\begin{table*}[t]
\centering
\caption{Qualitative regimes of network adaptation as a function of the reinforcement exponent $\mu$. Each regime corresponds to a distinct balance between efficiency, robustness, and decisiveness in Physarum-like dynamics.}
\label{tab:gamma_regimes}
\renewcommand{\arraystretch}{1.25}
\begin{tabularx}{\textwidth}{
    c@{\hspace{1em}}
    >{\raggedright\arraybackslash}X 
    >{\raggedright\arraybackslash}X 
    >{\raggedright\arraybackslash}X 
    >{\raggedright\arraybackslash}X
}
\toprule
\textbf{$\boldsymbol{\mu}$} & 
\textbf{Reinforcement} & 
\textbf{Morphology} & 
\textbf{Dynamics} & 
\textbf{Function} \\
\midrule

$\mu < 1$ &
Sublinear reinforcement &
Loopy, redundant network architecture &
Smooth, damped dynamics with multiple stable paths &
Exploratory behavior, robustness to perturbations \\

$\mu = 1$ &
Linear reinforcement &
Minimal, tree-like topology &
Gradient descent-like adaptation &
Balanced efficiency and stability \\

$\mu > 1$ &
Superlinear reinforcement &
Sparse, single-path connectivity &
Competitive dynamics with fast pruning &
Decisive responses, but structural brittleness \\

\bottomrule
\end{tabularx}
\end{table*}

\section{Physarum as a network}
\label{sec:physarum_network}

In this paper, we follow the theoretical description of \textit{Physarum} by the mathematical model by Tero, Kobayashi, and Nakagani (hereafter the TKN model), where the dynamics of \textit{Physarum} in a maze are mapped onto a network (\subfigref[c]{Physlagr2}).
The TKN considers the problem as described by the dynamics of tubes \cite{Tero2007} between different nodes (squares in \subfigref[c]{Physlagr2}) connected by links.
In all these experiments, food sources (FS, \subfigref[b--c]{Physlagr2}) are introduced.

Consider this connected graph $G = (V, E)$ representing a maze or some other set of external boundary conditions associated with a given problem.
Each undirected edge $e = (i, j)\in E$ has geometric length $L_{ij} > 0$ and a time-dependent \emph{hydraulic} conductance $D_{ij}(t) > 0$, defined as:
\begin{equation}
    D_{ij} = \left( \frac{\pi \, a_{ij}^4}{8 \, \cal{K}} \right).
\end{equation}
This includes the radius $a_{ij}$ of the section for the tube connecting two nodes, whereas $\cal{K}$ indicates the viscosity of the internal medium (roughly speaking, a measure of its resistance to deformation).
The intuitive meaning of $D_{ij}$ is a weight between the node pair $\{i, j\}$ that measures the efficiency of the flow.
The correlations between nodes suggest a Hebbian-like mechanism of adaptation, as will be discussed below.

Let $p_i$ denote the pressure at node $i$, and $Q_{ij}=-Q_{ji}$ the net flux from $i$ to $j$.
With source/sink injections $S_i$ satisfying $\sum_{i\in V} S_i=0$ (e.g., one source node with $+I$ and one sink with $-I$), the fast steady-state on the graph obeys the Ohm/Poiseuille law and Kirchhoff conservation, namely:
\begin{equation}
    Q_{ij} = \frac{D_{ij}}{L_{ij}} \, (p_i - p_j)
    \label{eq:ohm}
\end{equation}
In its standard form, this equation gives the pressure drop in an incompressible and Newtonian fluid in laminar flow through a long cylindrical pipe, where the cross-section is considered constant.
In our context, this constancy is an oversimplification, and the model description will require some dynamical feedback that changes the tube properties.

Similarly, we have the node conservation
\begin{equation}
    \sum_{j:(i, j)\in E} Q_{ij} = S_i
    \label{eq:kirchhoff}
\end{equation}
with $i = 1, 2, \dots, n$.

Tero \textit{et al.}~\cite{Tero2007} introduced a slow adaptation dynamics for the conductance of the generic form
\begin{equation}
    \frac{\mathrm{d}D_{ij}}{\mathrm{d}t} = f\!\left(|Q_{ij}|\right) - \delta \, D_{ij}.
    \label{eq:tero}
\end{equation}
This equation couples morphology to use: frequently used edges are reinforced, while underused edges decay.
The functional term is introduced by means of a scaling $f\!(x)\sim x^\mu$, where the exponent $\mu>0$ determines the nonlinearity of the reinforcement and $\delta>0$ is a decay rate.
In steady state,
\begin{equation}
    D_{ij} = \frac{1}{\delta}|Q_{ij}|^{\mu}.
\end{equation}
Combining this with equation~\eqref{eq:ohm} yields
\begin{equation}
    D_{ij}^{1-\mu} \propto |p_i - p_j|^{\mu} \, L_{ij}^{-\mu}.
\end{equation}
Three main cases can be considered here regarding the values of $\mu$: (a) linear reinforcement ($\mu = 1$), where conductance scales linearly with flow, leading to shortest-path structures; (b) superlinear reinforcement ($\mu>1$) can favor sparse, tree-like networks; and (c) sublinear reinforcement ($\mu<1$) allows for multiple paths to be maintained, yielding more redundant networks.

To characterize the extremal properties of \textit{Physarum}'s morphological computation, we define the \emph{\textit{Physarum} action} as the time integral of its Lagrangian:
\begin{equation}
    S_\phi[Q(t), \lambda(t); D(t)] = \int_{t_1}^{t_2} \mathcal{L}_\phi(Q(t), \lambda(t); D(t)) \, \mathrm{d}t.
    \label{eq:action}
\end{equation}
The stationarity of this functional, $\delta S_\phi = 0$, with respect to the fluxes $Q_{ij}(t)$ and the potentials $\lambda_i(t)$, produces the instantaneous Poiseuille--Kirchhoff equations governing steady flow within the network.

In the present framework, the term ``least action'' refers specifically to the \emph{instantaneous} minimization of the Dirichlet (or Thomson) functional by the fast hydrodynamic and potential variables.
In contrast, the slow morphological variables, denoted by \( D(t) \), evolve according to a dissipative process described as a gradient flow of a reduced free-energy functional.
This adaptation law for \( D(t) \) is not obtained from an Euler--Lagrange extremum of the action; rather, it follows an Onsager-type relaxation that monotonically decreases the free energy of the system.
Therefore, a fully time-dependent least-action principle for \( D(t) \) is beyond the scope of the present study.

\section{A Lagrangian for the steady network}
\label{sec:lagrangian_steady_network}

To remove the ambiguity associated with the potential's reference level, we fix the gauge by grounding one node, setting \( p_1 = 0 \).
We also restrict attention to connected graphs where all conductances satisfy \( D_{ij} > 0 \).
Under these conditions, the energy functional \( E(p; D) \) is strictly convex, guaranteeing the existence and uniqueness of the stationary potential configuration.

If the potential gauge is left unspecified, or if some conductances vanish within the interior of the feasible domain, the network Laplacian becomes singular.
In such cases, the stationary equations may not uniquely determine the node potentials.
To avoid this degeneracy, edges with vanishing conductance are treated as boundary elements that effectively prune themselves from the network.

\begin{figure*}[t]
    \centering
    \includegraphics[width=0.95\textwidth]{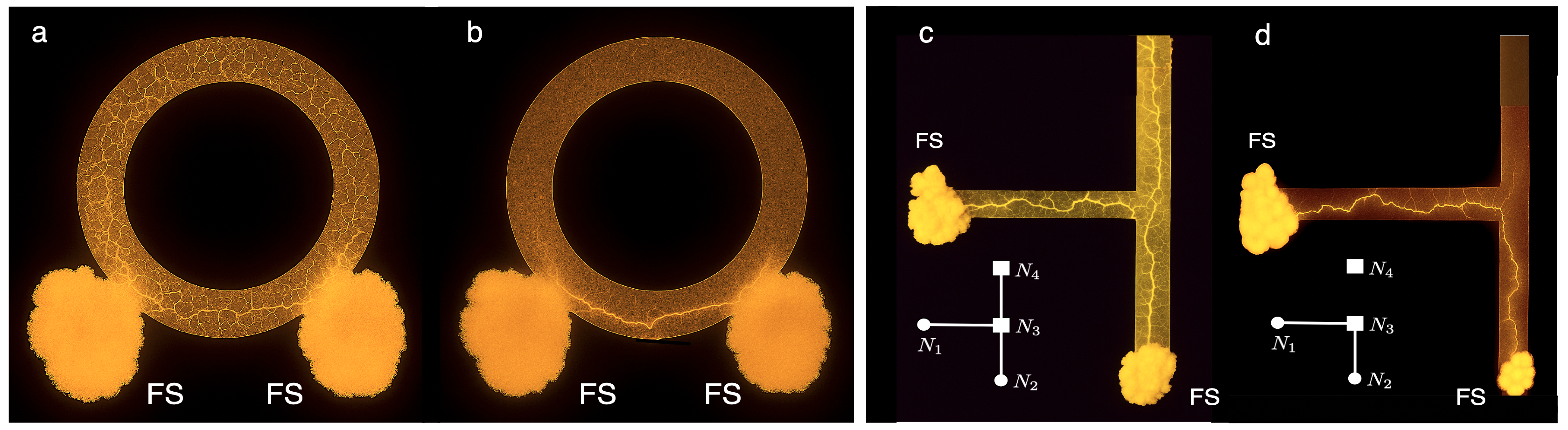}
    \caption{Shortest paths by \textit{Physarum} on simple graphs.
        Here, (a) and (b) indicate the initial and final state, respectively, of the ring-shaped network experiment.
        The lengths of longer and shorter paths are $L_1 = 42$ and $L_2 = 13$ mm, respectively.
        In (c) and (d), we display the initial and final states, respectively, of the T-shaped graph experiment.
        Images a--d redrawn using ChatGPT (GPT-5, OpenAI, 2025) based on the original photographs from~\cite{Tero2007}.
        The insets show, respectively, the initial and the final state of the simulation, which consisted of four nodes and three edges.
        The width of the black lines reflects the conductivity of each path.
    }
    \label{Physlagr3}
\end{figure*}

For \emph{fixed} conductances $D = \{ D_{ij} \}$, the steady flow--potential pair $(Q, p)$ solves Equations~\eqref{eq:ohm}--\eqref{eq:kirchhoff}.
Using the total power dissipated by viscous friction in all tubes, namely
\begin{equation}
    P(Q;D) = \sum_{(i, j)\in E} \frac{L_{ij}}{D_{ij}} \, Q_{ij}^2.
\end{equation}
the \textit{Physarum} Lagrangian $\mathcal{L}_{\phi} = \mathcal{L}_{\phi}(Q, \lambda;D)$ will be defined as
\begin{equation}
    \mathcal{L}_{\phi} = \frac{1}{2} \, \sum_{(i, j)\in E} \frac{L_{ij}}{D_{ij}} \, Q_{ij}^2 + \sum_{i\in V} \lambda_i \, \left( S_i - \sum_{\mathclap{j:(i, j)\in E}} Q_{ij} \right),
    \label{eq:Lagrange}
\end{equation}
where the first term is the instantaneous dissipated power, and the node multipliers $\lambda_i$ enforce conservation.
The corresponding Euler--Lagrange equations are
\begin{equation}
    \frac{\partial \mathcal{L}_{\phi}}{\partial Q_{ij}} = 0
\end{equation}
which gives the condition:
\begin{equation}
    \frac{L_{ij}}{D_{ij}} \, Q_{ij} - (\lambda_i - \lambda_j) = 0
    \implies
    Q_{ij} = \frac{D_{ij}}{L_{ij}} \, (\lambda_i - \lambda_j),
\end{equation}
and we also have:
\begin{equation}
\frac{\partial \mathcal{L}_{\phi}}{\partial \lambda_i}=0
\end{equation}
which gives $\sum_j Q_{ij} = S_i$.
Identifying $\lambda_i \equiv p_i$ we recover Equations~\eqref{eq:ohm}--\eqref{eq:kirchhoff}.
Eliminating $Q$ yields the dual principle \emph{Dirichlet/Thomson} on node potentials,
\begin{equation}
    \mathcal{E}(p; D) = \frac{1}{2} \, \sum_{(i, j)\in E} \frac{D_{ij}}{L_{ij}} \, (p_i - p_j)^2 - \sum_{i\in V} S_i \, p_i,
\label{eq:dirichlet}
\end{equation}
which we minimize in $p$ to get \eqref{eq:kirchhoff}.
At the minimizer $p^\ast(D)$, the dissipated power equals
\begin{gather}\begin{aligned}
    P(D) &= \sum_{(i, j)} \frac{L_{ij}}{D_{ij}} \, \left( Q_{ij}^{\ast} \right)^2 = \sum_i S_i \, p_i^\ast
    \\
    &= \min_{p} \sum_{(i, j)} \frac{D_{ij}}{L_{ij}} \, (p_i - p_j)^2,
    \label{eq:power}
\end{aligned}\end{gather}
a convex function of the conductances via the Rayleigh monotonicity of resistor networks.

To close the \emph{slow} morphology dynamics, introduce a reduced free-energy functional of the conductances that trades off (i) the power required to sustain the imposed fluxes with (ii) a metabolic maintenance cost proportional to tube length:
\begin{equation}
    \mathcal{F}(D) = P(D) + \beta \, \sum_{\mathclap{(i, j)\in E}} L_{ij} \, \Phi(D_{ij}),
\label{eq:F}
\end{equation}
where $\beta > 0$ weighs the maintenance cost relative to transport efficiency, and $\Phi$ is a convex per-length cost (e.g.,
$\Phi(D) = c \, D$ or $\Phi(D) = D^2/2$).
A natural dissipative evolution for $D$ is the gradient flow
\begin{equation}
    \frac{\mathrm{d} D_{ij}}{\mathrm{d}t} = -M_{ij}(D) \, \frac{\partial \mathcal{F}}{\partial D_{ij}},
\label{eq:gradientflow}
\end{equation}
with $M_{ij}(D)>0$, which guarantees $\dot{\mathcal{F}} \le 0$.
Using the envelope (Danskin) theorem and the optimal flows $Q^\ast(D)$ from \eqref{eq:Lagrange},
\begin{equation}
    \frac{\partial \cal{F}}{\partial D_{ij}} = -\frac{L_{ij}}{2 \, D_{ij}^2} \, \left( Q_{ij}^{\ast} \right)^2 + \beta \, L_{ij} \, \Phi(D_{ij}).
\label{eq:dF}
\end{equation}
Equation \eqref{eq:gradientflow} with \eqref{eq:dF} reproduces the phenomenology of Tero's law \eqref{eq:tero} for suitable choices.
For example, with mobility $M_{ij} = 2 \, D_{ij}^2/L_{ij}$ and $\Phi(D) = \delta \, D$ (linear maintenance),
\begin{equation}
    \frac{\mathrm{d} D_{ij}}{\mathrm{d}t} = \left( Q_{ij}^{\ast} \right)^2 - 2 \, \beta \, \delta \, D_{ij},
\end{equation}
a quadratic-stimulus variant.
Choosing instead a \emph{one-homogeneous} dissipation (a $p$-Laplacian variant) in the network part and $\Phi$ such that $\beta \, L_{ij} \, \Phi(D) \equiv 1$ reproduces the \emph{linear} stimulus $\dot{D_{ij}} \propto \left| Q_{ij}^\ast \right| - \delta \, D_{ij}$ used in \cite{Tero2007} (details are omitted here for space but straightforwardly replaced by the quadratic power in \eqref{eq:Lagrange} with an $L^1$-type term).

In all cases, steady states satisfy the first-order conditions
\begin{equation}
    \frac{\partial \mathcal{F}}{\partial D_{ij}} = 0
    \Longleftrightarrow
    \frac{\left( Q_{ij}^{\ast} \right)^2}{D_{ij}^2} = 2 \, \beta \, D_{ij} \, \Phi(D_{ij}),
    \label{eq:stationaryD}
\end{equation}
which relate optimal morphology to usage (flux).
In the following sections, we apply these formal results to three case studies that illustrate how the proposed framework describes the steady-state dynamics of the slime mold on graphs as an outcome of constrained minimization of energy dissipation.

\section{The circular ring problem}
\label{sec:ring}

Solving the general problem of \textit{Physarum} within a complex set of boundaries is not an easy task, but an elegant and illustrative example is provided by the ring case study, as shown in \subfigref[a--b]{Physlagr3}~\cite{nakagaki2001path, Tero2007}.
In mathematical terms, we study the two-dimensional system derived as a special case of the TKN model:
\begin{equation}
    \frac{\mathrm{d}D_i}{\mathrm{d}t} = \left(
        \frac{\dfrac{D_i}{L_i}}{\dfrac{D_1}{L_1} + \dfrac{D_2}{L_2}}
    \right)^{\mu} - \delta \, D_i,
\end{equation}
with $i=1, 2$.

Let us now show how the \textit{Physarum} variational formulation for this ring-shaped network recovers the same results as the dynamical equations.

The system contains two terminal nodes: a source $s$ injecting a constant flow $+I$, and a sink $t$ extracting the same amount $-I$.
The two nodes are connected by two distinct arcs, $a$ and $b$, forming two parallel pathways between $s$ and $t$.
Each arc represents an effective tubular branch of the plasmodium, characterized by its length $L_j$ and its hydraulic conductance $D_j$, with $j\in\{a, b\}$.

Let $p_s$ and $p_t$ denote the pressures at the source and sink, respectively, and define the potential drop across the ring as $\Delta p = p_s - p_t$.
Both arcs experience the same pressure difference, as they connect the same terminals.
According to Poiseuille's law, the flow through each arc is proportional to the pressure drop and its conductance-to-length ratio:
\begin{align}
    Q_j &= \frac{D_j}{L_j}\, \Delta p,
    &
    j &\in \{ a, b \}.
\end{align}
At the source node, Kirchhoff's law requires that the total outflow equals the imposed injection $I$, i.e.,
\begin{equation}
    Q_a + Q_b = I.
\end{equation}
These two relations fully describe the stationary hydrodynamic state for fixed conductances $D_a$, $D_b$, and imposed current $I$.

The same relations can be derived variationally using the \textit{Physarum} Lagrangian, which clarifies the interplay between energy dissipation and conservation constraints.
For the two-node ring, the independent flows are $\{ Q_a, Q_b \}$, and there is only one relevant conservation constraint at the source: $S_s - (Q_a + Q_b) = 0$ with $S_s = +I$.
Identifying $\lambda_s = p_s$ and $\lambda_t = p_t$, we can write the Lagrangian as
\begin{widetext}
\begin{gather}\begin{aligned}
    \mathcal{L}_\phi(Q_a, Q_b, p_s, p_t; D_a, D_b)
    &= \frac{1}{2} \, \left( \frac{L_a}{D_a} \, Q_a^2 + \frac{L_b}{D_b} \, Q_b^2 \right) + p_s \, (I - Q_a - Q_b) + p_t \, (-I + Q_a + Q_b)
    \\
    &= \frac{1}{2} \, \left( \frac{L_a}{D_a} \, Q_a^2 + \frac{L_b}{D_b} \, Q_b^2 \right) - (p_s - p_t) \, (Q_a + Q_b - I).
    \label{eq:ringL}
\end{aligned}\end{gather}
\end{widetext}
The two parts of Equation~\eqref{eq:ringL} have distinct physical meanings.
The quadratic term measures the energy dissipated by viscous friction in both arcs and thus penalizes large flows through high-resistance channels.
The second term imposes global mass conservation: deviations of the total flow $(Q_a + Q_b)$ from the imposed value $I$ are penalized through the multiplier $(p_s - p_t)$, which will emerge as the pressure drop $\Delta p$.

The steady-state relations follow from the stationary conditions of $\mathcal{L}_\phi$.
Taking derivatives with respect to $Q_a$ and $Q_b$ gives
\begin{align}
    \frac{\partial \mathcal{L}_\phi}{\partial Q_a} &= \frac{L_a}{D_a} \, Q_a - (p_s - p_t) = 0,
    \\
    \frac{\partial \mathcal{L}_\phi}{\partial Q_b} &= \frac{L_b}{D_b} \, Q_b - (p_s - p_t) = 0,
\end{align}
which yields
\begin{align}
    Q_a &= \frac{D_a}{L_a}\Delta p,
    &
    Q_b &= \frac{D_b}{L_b}\Delta p.
\end{align}

Variation with respect to $(\Delta p = p_s - p_t)$ yields
\begin{equation}
    \frac{\partial \mathcal{L}_\phi}{\partial(\Delta p)} = -(Q_a + Q_b - I) = 0,
\end{equation}
recovering Kirchhoff's law $Q_a + Q_b = I$.
The stationary point of the Lagrangian therefore reproduces the Poiseuille and conservation equations.

To eliminate the flows and obtain a reduced form depending only on $\Delta p$, we substitute these relations into Equation~\eqref{eq:ringL}, obtaining
\begin{widetext}
\begin{gather}\begin{aligned}
    \mathcal{L}_\phi(\Delta p; D_a, D_b) &= \frac{1}{2} \, \left[ 
        \frac{L_a}{D_a} \left( \frac{D_a}{L_a} \, \Delta p \right)^2 + \frac{L_b}{D_b} \, \left( \frac{D_b}{L_b} \, \Delta p \right)^2
    \right] - \Delta p \, \left( Q_a + Q_b - I \right)
    \\
    &= \frac{1}{2} \, \left( \frac{D_a}{L_a} + \frac{D_b}{L_b} \right) \, \Delta p^2 - \Delta p \, \left[
        \left( \frac{D_a}{L_a} + \frac{D_b}{L_b} \right) \, \Delta p - I
    \right].
\end{aligned}\end{gather}
\end{widetext}

Stationarity with respect to $\Delta p$ yields
\begin{equation}
    \frac{\partial \mathcal{L}_\phi}{\partial(\Delta p)} = 0
    \implies
    \left(\frac{D_a}{L_a}+\frac{D_b}{L_b}\right)\Delta p = I,
\end{equation}
so that
\begin{equation}
    \Delta p = \frac{I}{D_a/L_a + D_b/L_b}.
\end{equation}
Substituting this equilibrium condition back into $\mathcal{L}_\phi$ gives the minimized Lagrangian
\begin{equation}
    \mathcal{L}_\phi^\ast(D_a, D_b) = \frac{I^2}{D_a/L_a + D_b/L_b}.
\end{equation}
This quantity equals the total power dissipated in the ring for fixed $D_a$, $D_b$, and current $I$.
The denominator represents the sum of the two effective conductances of the arcs, so $\mathcal{L}_\phi^\ast = I^2 \, R_\text{eq}$, with $R_\text{eq}$ the equivalent resistance.
The variational formalism thus reproduces the standard circuit result, but with a structure that naturally accommodates dynamic adaptation.

To account for the metabolic cost of maintaining tube volume, one introduces a per-length maintenance term $\Phi(D) = \delta \, D$ with weight $\beta$.
The total ``free energy'' becomes
\begin{equation}
    \mathcal{F}(D_a, D_b) = \frac{I^2}{D_a/L_a + D_b/L_b} + \beta \, \delta \, (L_a \, D_a + L_b \, D_b).
\end{equation}

Minimizing $\mathcal{F}$ with respect to the conductances yields
\begin{align}
    \frac{\partial \mathcal{F}}{\partial D_a} &= -\frac{I^2}{(D_a/L_a + D_b/L_b)^2} \, \frac{1}{L_a} + \beta \, \delta \, L_a = 0,
    \\
    \frac{\partial \mathcal{F}}{\partial D_b} &= -\frac{I^2}{(D_a/L_a + D_b/L_b)^2} \, \frac{1}{L_b} + \beta \, \delta \, L_b = 0.
\end{align}
Both conditions can hold simultaneously only if $L_a = L_b$.
If the arcs have unequal lengths, no interior critical point exists, and the minimum lies on the boundary where one conductance vanishes.
If $L_a < L_b$, the minimum occurs at $D_b = 0$,
\begin{equation}
    \mathcal{F}(D_a, 0) = \frac{I^2 L_a}{D_a} + \beta \, \delta \, L_a \, D_a,
\end{equation}
whose derivative with respect to $D_a$ yields
\begin{equation}
    -\frac{I^2 \, L_a}{D_a^2} + \beta \, \delta \, L_a = 0
    \implies
    D_a^\star = \frac{I}{\sqrt{\beta \, \delta}}.
\end{equation}
The corresponding minimum is
\begin{equation}
    \mathcal{F}_{\min | b\equiv 0} = 2 \, I \, \sqrt{\beta \, \delta} \, L_a.
\end{equation}
A symmetric argument applies if $L_b < L_a$.
The system therefore selects the shorter path, setting the conductance of the longer one to zero.
The \textit{Physarum} network thus prunes the less efficient arc and retains the shortest route, reproducing the empirically observed behavior of the organism when solving a maze.

The ring model explicitly shows that the \textit{Physarum} Lagrangian encodes both the flow physics and the energetics of dissipation.
Eliminating flows leads to the total power~$\mathcal{L}_\phi^\ast$, while adding the maintenance cost yields a free-energy landscape~$\mathcal{F}(D_a, D_b)$ whose minimum corresponds to the optimal morphology observed.
This simple two-branch system captures the essence of Physarum's adaptive computation: a continuous competition between parallel paths that converges toward the least-dissipation configuration.
The resulting structure arises purely from physical principles: energy minimization and feedback, without any external control, demonstrating how a simple living material implements an analog form of decision-making through least-dissipation dynamics: instantaneous Dirichlet minimization for flows and an Onsagerian gradient flow for morphology.

\section{The two-branch problem}
\label{sec:twobranch}

Many experiments on biological and physical decision making are designed around a two-branch, or binary, choice structure.
This simplification serves several purposes.
First, it isolates the fundamental mechanism of selection between competing alternatives while minimizing confounding variables.
A two-branch setup allows researchers to quantify preference, response time, or efficiency in the clearest possible way, since the outcome can be expressed as a direct comparison between two options.
Second, many natural decision processes, from chemotactic movement to neuronal activation or foraging, can be reduced to bistable dynamics, where feedback mechanisms amplify one branch while suppressing the other.
Finally, the two-branch model provides a convenient starting point for theoretical analysis: it corresponds to the simplest nontrivial case of symmetry breaking, which can later be generalized to multiple competing pathways or continuous decision spaces \cite{sole2011phase}.

The two-branch experiment for \textit{Physarum} as presented in \cite{tero2010rules} is displayed in \subfigref[c--d]{Physlagr3}.
The food sources are located at two extremes of the channels, but are missing in the third one.
We can formalize this as a simple bifurcating network in which a source node $a$ feeds two downstream nodes $b$ and $c$ through an intermediate junction $z$ (see also the insets in \subfigref[c--d]{Physlagr3}).
The geometry of the problem is thus given by:
\[
    a \longrightarrow z \longrightarrow
    \begin{cases}
        b \ (\text{sink, } S_b = -I),
        \\
        c \ (\text{dead end, } S_c = 0),
    \end{cases}
\]
with $S_a = +I$ and $S_z = 0$.
Each edge $(i, j)$ has a length $L_{ij}$, conductance $D_{ij}$, and flux $Q_{ij}$ defined positive along the flow direction.

For this specific graph, with edges $(a, z)$, $(z, b)$, $(z, c)$, the Lagrangian becomes
\begin{widetext}
\begin{align}
    \mathcal{L}_\phi &= \frac{1}{2} \, \left(
         \frac{L_{az}}{D_{az}} \, Q_{az}^2
        +\frac{L_{zb}}{D_{zb}} \, Q_{zb}^2
        +\frac{L_{zc}}{D_{zc}} \, Q_{zc}^2
    \right)
    + p_a \, (I - Q_{az})
    + p_z \, (-Q_{zb} - Q_{zc} + Q_{az})
    + p_b \, (-I + Q_{zb})
    + p_c \, (Q_{zc}).
    \label{eq:Y_L_explicit}
\end{align}
\end{widetext}

Stationarity with respect to the flows gives the Poiseuille relations:
\begin{align}
    \frac{\partial \mathcal{L}_\phi}{\partial Q_{az}}=0
    &\implies \frac{L_{az}}{D_{az}}Q_{az} = p_a - p_z,
    \\[0.5em]
    \frac{\partial \mathcal{L}_\phi}{\partial Q_{zb}}=0
    &\implies \frac{L_{zb}}{D_{zb}}Q_{zb} = p_z - p_b,
    \\[0.5em]
    \frac{\partial \mathcal{L}_\phi}{\partial Q_{zc}}=0
    &\implies \frac{L_{zc}}{D_{zc}}Q_{zc} = p_z - p_c.
\end{align}
Stationarity with respect to the node pressures enforces conservation:
\begin{align}
    \frac{\partial \mathcal{L}_\phi}{\partial p_a} &: I - Q_{az} = 0,
    \\[0.5em]
    \frac{\partial \mathcal{L}_\phi}{\partial p_z} &: Q_{az} - Q_{zb} - Q_{zc} = 0,
    \\[0.5em]
    \frac{\partial \mathcal{L}_\phi}{\partial p_b} &: Q_{zb} - I = 0,
    \\[0.5em]
    \frac{\partial \mathcal{L}_\phi}{\partial p_c} &: Q_{zc} = 0.
\end{align}
From these relations:
\[
    Q_{az} = Q_{zb} = I, \qquad Q_{zc} = 0.
\]
Thus, the lateral branch $z\!\to\!c$ carries no flow.
Since the conductances evolve as $\dot D_{ij} = |Q_{ij}| - D_{ij}$, we obtain
\[
    \frac{\mathrm{d}D_{zc}}{\mathrm{d}t}= -D_{zc},
\]
implying exponential decay $D_{zc}(t) \to 0$.

The total dissipated power is
\begin{equation}
    P = \frac{L_{az}}{D_{az}} \, Q_{az}^2
      + \frac{L_{zb}}{D_{zb}} \, Q_{zb}^2
      + \frac{L_{zc}}{D_{zc}} \, Q_{zc}^2.
\end{equation}
Minimizing $P$ subject to the above constraints enforces $Q_{zc}=0$ because any nonzero flow in the dead-end branch would increase dissipation without contributing to transport between source and sink.
Therefore, the variational principle predicts spontaneous pruning of the unused branch, leaving a single efficient path connecting $a$ to $b$.

\begin{figure}
    \centering
    \includegraphics[width=8 cm]{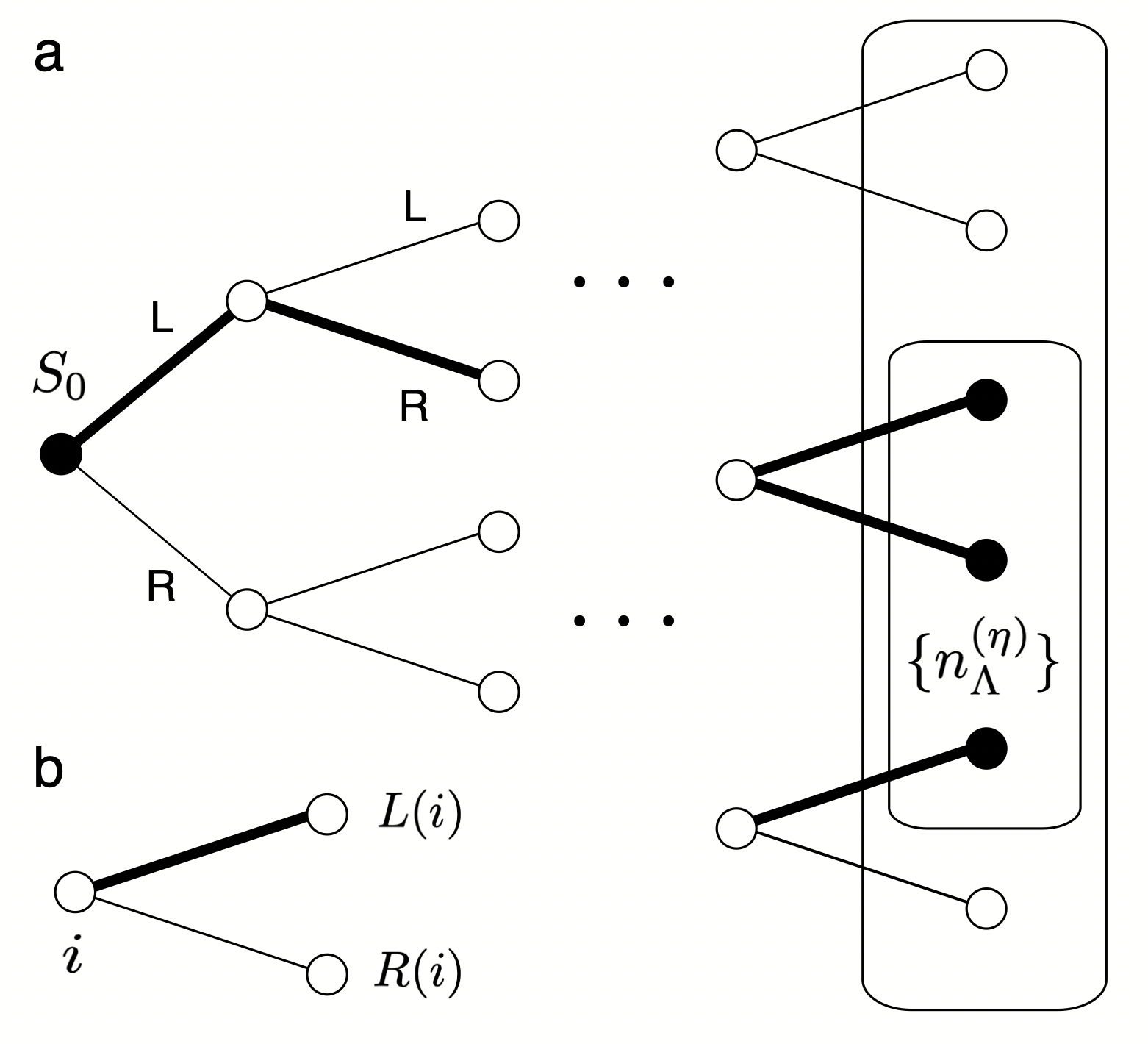}
    \caption{Schematic representation of the binary tree topology used in the \textit{Physarum} variational model.
        In (a), the root node (black circle on the left) represents the nutrient source, and each internal node branches into a left (\(L\)) and a right (\(R\)) descendant.
        At depth \(\Lambda\), some terminal nodes (black circles on the right) act as sinks, while others remain inactive.
        The network fluxes \(Q_{iL}\) and \(Q_{iR}\) evolve according to the Poiseuille relations, and the conductances \(D_{ij}\) adapt dynamically, leading to the pruning of all branches that do not connect the source to at least one active sink.
        In (b), the notation used for the left/right nodes is indicated.
    }
    \label{Physlagr1}
\end{figure}

Although the paper presenting the application of the TKN model only considers this simple tree, our formalism can be extended to explore a more complex scenario.
The problem can be generalised to a tree with the source at the root and choosing one of the tips of the tree (its leaves) as the potential sinks (see \subfigref{Physlagr1}).
To model this problem, we consider a binary rooted tree of depth $\Lambda$ with node set $V$ and edge set $E$.
The \textit{root} node acts as a source, while a subset of the leaves $\{n_\Lambda^{(\eta)}\}$ are sinks with fixed extraction rates $I_\eta>0$.
All other leaves are inactive.
Thus, $S_0 = \sum_{\eta} I_\eta$ and $S_{n_\Lambda^{(\eta)}} = -I_\eta$ while  
$S_i = 0$ for all other nodes in  $V$ so that $\sum_i S_i = 0$.

Following our previous notation, each edge $(i,j)\in E$ has length $L_{ij}>0$, conductance $D_{ij}>0$, and scalar flux $Q_{ij}$, defined positive in the direction from $i$ to $j$, and each node $i$ carries a pressure variable $p_i$.
For every internal node $i$, we denote by $L(i)$ and $R(i)$ its left and right descendants, respectively (\subfigref[b]{Physlagr1}).
The sets
\begin{align*}
    \mathrm{Out}(i) &= \left\{ L(i), R(i) \right\},
    &
    \mathrm{In}(i) &= \{ \text{parent}(i) \},
\end{align*}
specify the directed structure.

The Lagrangian functional governing the stationary flow configuration is
\begin{gather}\begin{aligned}
    \mathcal{L}_\phi &= \frac{1}{2} \, \sum_{(i,j)\in E} \frac{L_{ij}}{D_{ij}} \, Q_{ij}^2
    \\
    &+ \sum_{i\in V} p_i \, \left(
        S_i + \sum_{\mathclap{j\in \mathrm{In}(i)}} Q_{ji} - \sum_{\mathclap{k\in \mathrm{Out}(i)}} Q_{ik}
    \right).
\label{eq:L_full}
\end{aligned}\end{gather}
where the first term represents total dissipation and the second enforces local conservation of flow.
Stationarity with respect to $Q_{ij}$ and $p_i$ gives familiar expressions:
\begin{align*}
    \frac{L_{ij} \, Q_{ij}}{D_{ij}} &= p_i - p_j,
    &
    S_i + \sum_{\mathclap{j\in \mathrm{In}(i)}} Q_{ji} - \sum_{\mathclap{k\in \mathrm{Out}(i)}} Q_{ik} &= 0.
\end{align*}
For a binary branching node $i$, the local contribution to the 
Lagrangian is
\begin{gather}\begin{aligned}
    \mathcal{L}_\phi^{(i)} &= \frac{1}{2} \, \left(
        \frac{L_{iL}}{D_{iL}} \, Q_{iL}^2 + \frac{L_{iR}}{D_{iR}} \, Q_{iR}^2
    \right)
    \\
    &+ p_i \, \left(
        S_i + Q_i^{\mathrm{in}} - Q_{iL} - Q_{iR}
    \right),
\label{eq:Llocal_finalcompact}
\end{aligned}\end{gather}
where $Q_i^{\mathrm{in}}$ denotes the inflow from the parent node~$i$.
Minimizing~\eqref{eq:Llocal_finalcompact} with respect to $Q_{iL}$ and $Q_{iR}$, under the constraint $Q_i^{\mathrm{in}} = Q_{iL} + Q_{iR}$, gives
\begin{gather}\begin{aligned}
    Q_{iL} &= Q_i^{\mathrm{in}} \, \frac{G_{iL}}{G_{iL} + G_{iR}},
    \\[0.5em]
    Q_{iR} &= Q_i^{\mathrm{in}} \, \frac{G_{iR}}{G_{iL} + G_{iR}},
    \label{eq:Qpartition_finalcompact}
\end{aligned}\end{gather}
where $G_{i\alpha} = D_{i\alpha}/L_{i\alpha}$, $\alpha = \{ L, R \}$, denotes the effective branch conductance per unit length.

To determine which branches remain active, consider the subtree $T(i)$ rooted at node~$i$.
If no sink lies within $T(i)$, i.e., $S_j=0$ for all $j \in  T(i)$, summing over $T(i)$ gives
\begin{equation}
    \sum_{j\in T(i)} S_j = Q_{\mathrm{parent}(i),i} - \sum_{\mathclap{k \in \mathrm{Out}(i)}} Q_{ik} = 0,
\end{equation}
which implies $Q_{\mathrm{parent}(i), i} = 0$ and consequently $Q_{ik}=0$ for every descendant edge $(i, k) \in T(i)$.
Thus, any subtree without sinks carries no flux.
Applying this argument recursively from the leaves upward prunes all inactive branches.

Define for each internal node $i$ the total downstream sink currents
\begin{gather}\begin{aligned}
    \Sigma_{iL} &:= \sum_{n_\Lambda^{(\eta)} \in T(L(i))} I_\eta,
    \\[0.5em]
    \Sigma_{iR} &:= \sum_{n_\Lambda^{(\eta)} \in T(R(i))} I_\eta.
\end{aligned}\end{gather}
Conservation gives $Q_i^{\mathrm{in}} = \Sigma_{iL} + \Sigma_{iR}$, and substituting into Equation~\eqref{eq:Qpartition_finalcompact} yields
\begin{gather}\begin{aligned}
    Q_{iL} &= \left( \Sigma_{iL} + \Sigma_{iR} \right) \, \frac{G_{iL}}{G_{iL} + G_{iR}},
    \\[0.5em]
    Q_{iR} &= \left( \Sigma_{iL} + \Sigma_{iR} \right) \, \frac{G_{iR}}{G_{iL} + G_{iR}}.
    \label{eq:Qdist_finalcompact}
\end{aligned}\end{gather}
If $\Sigma_{iL}=0$, then $Q_{iL}=0$, and similarly if $\Sigma_{iR}=0$, then 
$Q_{iR}=0$.
Hence, a branch carries a nonzero flux if and only if its subtree contains at least one sink.

Let $E_*$ denote the set of all active edges, i.e., those belonging to at least one source–sink path.
For each active edge $(i,j)\in E_*$,
\begin{align}
    Q_{ij} &= \sum_{n_\Lambda^{(\eta)} \in T(j)} I_\eta,
    &
    Q_{ij} &= 0 \text{ if } (i,j) \notin E_*.
\end{align}
The total power dissipation is
\begin{equation}
    P = \sum_{(i, j) \in E_*} \frac{L_{ij}}{D_{ij}} \, Q_{ij}^2,
    \label{eq:P_finalcompact}
\end{equation}
and any flow outside $E_*$ would increase $P$ without reducing 
source–sink potentials.
Thus, the minimal-dissipation solution of Equation~\eqref{eq:L_full} corresponds exactly to $E_*$.

Finally, the \textit{Physarum} conductance dynamics from the KTN model,  $\dot D_{ij}=|Q_{ij}|-D_{ij}$, drives the system toward this stationary configuration: along active edges, $D_{ij} \to |Q_{ij}| > 0$, while inactive ones decay as $D_{ij}(t) \sim e^{-t}$.
The long-time limit coincides with the minimizer of $\mathcal{L}_\phi$.
In summary, in a binary tree with one source and arbitrary sinks, the stationary solution of the \textit{Physarum} dynamics minimizes $\mathcal{L}_\phi$ under the conservation constraints.
At equilibrium, we have
\begin{gather}\begin{aligned}
    Q_{ij} &= 0, & \text{ for } & (i, j) \notin E_*
    \\
    D_{ij} &= |Q_{ij}|, & \text{ for } & (i, j) \in E_*,
\end{aligned}\end{gather}
where $E_*$ is the unique minimal set of edges that connect the source to all sinks.
Therefore, both the variational and dynamical formulations select the same minimal-dissipation transport tree, explaining why only the source--sink paths survive while all other branches decay.

\section{Square lattice}
\label{sec:lattice}

We now turn to a more complex and illustrative setting: a square lattice with food sources placed at two opposing external vertices---specifically, the top-left and bottom-right corners.
Consider an \(N \times N\) lattice graph \(G = (V, E)\) embedded in the unit square \(\Omega = [0, 1]^2\).
The lattice spacing is \(h = 1/(N-1)\), so node \((i, j)\) maps to the physical location
\begin{align*}
    x_i &= (i-1) \, h,
    &
    y_j &= (j-1) \,h,
\end{align*}
and each edge has length \( h \).

A current \(I\) is injected at the source \((1, 1)\) and extracted at the sink \((N, N)\), modeled by the discrete source term
\[
    S_{ij} = \begin{cases}
        \hfill +I, & i = j = 1,
        \\
        \hfill -I, & i = j = N,
        \\
        \hfill 0, & \text{otherwise}.
    \end{cases}
\]

Each edge in the graph, \( e = \{(i, j), (k, \ell)\} \in E \), carries a time-dependent conductance \(D_e(t) > 0\), representing the tube's transport capacity.
To simplify notation for local sums, we define the neighborhood of a node as the set of its von Neumann adjacent neighbors,
\[
    \mathcal{N}_{ij} = \left\{ (i \pm 1, j), (i, j \pm 1) \right\} \cap V,
\]
and the edge neighborhood as the set of edges incident to the node,
\[
    \mathcal{I}_{ij} = \left\{\{(i, j), (k, \ell)\} \in E : (k, \ell) \in \mathcal{N}_{ij} \right\}.
\]

The fast equilibria are derived from the discrete Lagrangian~\eqref{eq:Lagrange},
\begin{gather}\begin{aligned}
    \mathcal{L}_\phi(Q, \lambda; D) &= \frac{1}{2} \sum_{e \in E} \frac{h}{D_e} Q_e^2
    \\
    &+ \sum_{\mathclap{(i, j) \in V}} \lambda_{ij} \Bigl( S_{ij} - \sum_{e \in \mathcal{I}_{ij}} Q_e \Bigr),
\end{aligned}\end{gather}
where \(Q_e\) is the antisymmetric flow on edge \(e\).

Stationarity of \(\mathcal{L}_\phi\) with respect to the flow variables yields the discrete Ohm's law for an edge \( e \),
\begin{align}
    Q_e &= \frac{D_e}{h} \, (p_{ij} - p_{k\ell}), & p_{ij} &:= \lambda_{ij},
\label{eq:ohm_law_edge}
\end{align}
which states that flow is proportional to the pressure difference, with conductance scaled by the inverse edge length.
Stationarity with respect to the multipliers \(\lambda_{ij}\) enforces Kirchhoff's current law at each node,
\[
    \sum_{e \in \mathcal{I}_{ij}} Q_e = S_{ij},
\]
ensuring conservation of mass or charge.

Substituting Ohm's law into the conservation law eliminates the flows and gives a discrete elliptic equation for the pressure:
\begin{equation}
    \sum_{(k, \ell) \in \mathcal{N}_{ij}} \frac{D_e}{h} (p_{ij} - p_{k\ell}) = S_{ij}.
    \label{eq:discrete_conservation}
\end{equation}

In the continuum limit \( h \to 0^+ \), each interior node \((i,j)\) corresponds to a square control volume of area \( h^2 \).
Dividing both sides of \eqref{eq:discrete_conservation} by \( h^2 \) gives a balance of current densities:
\begin{equation}
    \frac{1}{h^2} \, \sum_{(k, \ell) \in \mathcal{N}(i,j)} \frac{D_e}{h} \, (p_{ij} - p_{k\ell}) = \frac{S_{ij}}{h^2}.
    \label{eq:discrete_conservation_scaled}
\end{equation}
The left-hand side converges to the divergence of the continuous flux \( \mathbf{J} = -D \, \nabla p \), i.e.,
\[
    \frac{1}{h^2} \sum_{(k, \ell) \in \mathcal{N}(i,j)} \frac{D_e}{h} \, (p_{ij} - p_{k\ell}) \to \nabla \cdot (D \, \nabla p).
\]
The right-hand side represents the source density, which, since the discrete sources are concentrated at individual nodes, the expression converges in the sense of distributions to a pair of Dirac delta functions:
\[
    \frac{S_{ij}}{h^2} \to I \, \big[ \delta(x) \, \delta(y) - \delta(x-1) \, \delta(y-1) \big].
\]

Consequently, in the continuum limit and for a smooth conductance field \( D \), the pressure \( p \) satisfies the elliptic equation
\begin{equation}
    \nabla \cdot (D \, \nabla p) = -I \, \big[ \delta(x) \, \delta(y) - \delta(x-1) \, \delta(y-1) \big].
    \label{eq:conservation_law}
\end{equation}

The conductance evolves slowly according to an approximate adaptation rule~\eqref{eq:tero},
\[
    \frac{\mathrm{d}D_e}{\mathrm{d}t} = |Q_e|^\mu - \delta \, D_e,
\]
where \(\mu > 0\) controls the nonlinearity of reinforcement and \(\delta > 0\) models decay due to lack of use.
For a smooth pressure field \(p\), \(|p_{ij} - p_{k\ell}| \approx h\,|\nabla p|\) in the edge direction.
Combined with the discrete Ohm's law~\eqref{eq:ohm_law_edge},
\[
    |Q_e| \approx D_e \, |\nabla p|.
\]  

Under a suitable rescaling of time, the discrete dynamics converge to the local continuum equation  
\begin{equation}
    \frac{\partial D}{\partial t} = \big( |\nabla p| \, D \big)^\mu - \delta \, D,
\label{eq:continuum_adaptation}
\end{equation}  
which reinforces conductance in regions of high flow while eliminating unused pathways.

For \(\mu = 1\), the adaptation dynamics are linear in flux magnitude, leading to an explicit steady state.
Setting \(\partial D / \partial t = 0\) in \eqref{eq:continuum_adaptation}, the steady state must satisfy
\[
    D \, \big( |\nabla p| - \delta \big) = 0.
\]
Because a nonzero current \(I\) is injected at the source and extracted at the sink, the conductance cannot vanish everywhere; it must be positive on at least one connecting path.
Consequently, the pressure gradient satisfies the eikonal equation  
\begin{equation}
    |\nabla p| = \delta.
\end{equation}
Combined with the conservation law \eqref{eq:conservation_law}, this implies that the flow is confined to curves along which the pressure drops at a constant rate \(\delta\).

Because the pressure difference between source and sink is fixed, only paths of minimal Euclidean length can sustain such a constant gradient without violating the single-valuedness of \(p\).
Consequently, in the continuum limit, the conductance concentrates on the unique shortest path connecting \((0, 0)\) to \((1, 1)\)---the straight diagonal \(y = x\).

\begin{figure}
    \centering
    \includegraphics[width=8.5 cm]{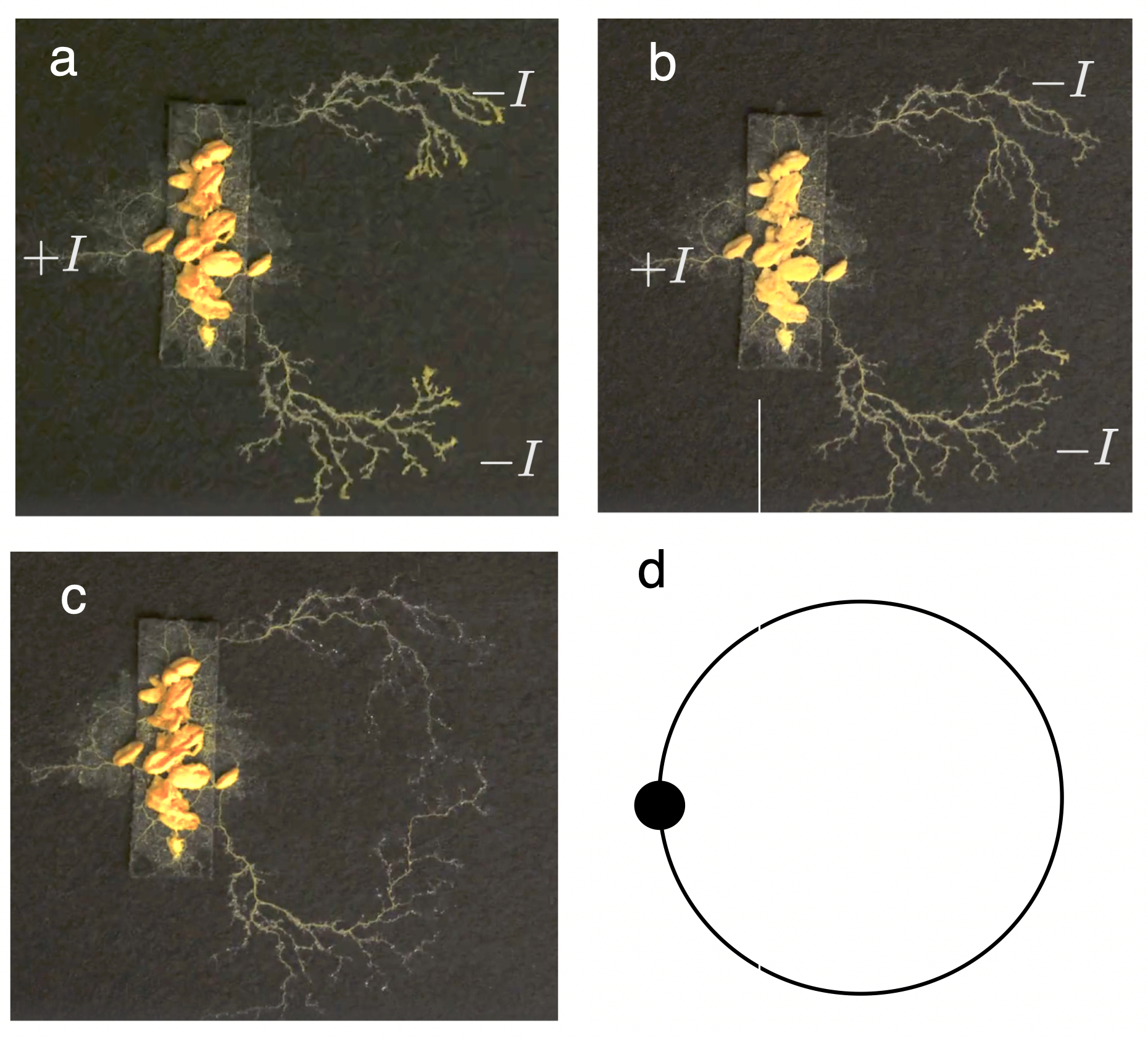}
    \caption{Formation and decay of a self-connected loop in \textit{Physarum polycephalum}.
        Panels (a–c) show a time sequence of the organism expanding from the central food source through two opposite branches, which eventually merge to form a closed loop (c).
        After contact, the entire network retracts and disappears, indicating loss of flow and nutrient transport.
        Panel (d) represents the corresponding idealized topology, where the source and sink coincide ($s = t$), forming a single closed cycle.
    }
    \label{fig:loop}
\end{figure}

\section{Discussion}
\label{sec:discussion}

The slime mould \textit{Physarum polycephalum} has long attracted interest across physics, biology, and cognitive science \cite{le2024physarum}.
Its adaptive morphology and rich dynamics enable memory \cite{boussard2019memory}, habituation \cite{boisseau2016habituation}, learning \cite{boussard2021adaptive, dussutour2021learning}, spatio-temporal oscillations \cite{takamatsu2001spatiotemporal}, and self-organized transport networks \cite{baumgarten2010plasmodial}.
These capabilities have inspired the field of ``\textit{Physarum} computation,'' where the organism performs network optimization tasks comparable to algorithmic problem solving \cite{adamatzky2012physarum, schumann2014bio, adamatzky2015atlas, bajpai2025morphological, liu2013physarum, awad2023survey}.

In this work, we have analyzed these behaviors through a physics perspective, showing that the adaptive dynamics of \textit{Physarum} can be captured within a Lagrangian framework.
This variational structure embodies a least-action principle, where the system organizes itself to minimize global dissipation under metabolic constraints.
The Lagrange functional introduced here encodes the balance between the energetic cost of maintaining the protoplasmic network and the benefit of efficient transport.
We have shown how to build this Lagrangian for three well-defined case studies, namely the ring, the tree graph, and a square lattice, showing that our energy dissipation under constraints fully captures the final states achieved by this aneural organism. At steady state, the morphology and flux distribution correspond to extrema of this Lagrangian, revealing that \textit{Physarum}'s network optimization emerges as a consequence of least action, not of explicit computation or sensory awareness.

This interpretation is directly supported by the loop-formation experiment shown in (\subfigref[a--c]{fig:loop}), where exploratory branches initially grow and approach each other, to eventually fuse, and then collapse.
Although the observed pattern suggests a sophisticated perception--reaction response to a poor environment, the mechanism is simple.
When the Lagrangian degenerates as a result of the formation of a closed loop (\subfigref[d]{fig:loop}) with no source--sink topology, the decay law $\dot D_{ij} = -D_{ij}$ follows, predicting the observed exponential vanishing of conductances.
Such behavior reflects local feedback and intrinsic physical constraints rather than higher-order planning.

Our results indicate that a physics-based principle, shared with well-known transport systems, offers a coherent picture of the behavior of \textit{Physarum}.
Once boundary conditions are specified, the organism exploits its own search dynamics and path-minimization capacity to solve transport problems.
In this view, the apparent ``intelligent'' behavior displayed under spatially constrained conditions does not reflect an intrinsic cognitive faculty but rather an emergent property revealed within an artificial setting.
Nevertheless, this perspective highlights the remarkable potential of a living system that, by harnessing active matter mechanisms, can display complex adaptive responses under uncertain environmental conditions.
Within the broader morphospace of multicellular and pre-neural cognition \cite{Olle2016Morphospace, sole2024open}, \textit{Physarum} represents a singular case study---an aneural organism whose least-action dynamics illuminate how physics-based principles can give rise to problem-solving behaviors usually associated with more complex systems.

\begin{acknowledgments}
    The authors thank the members of the Complex Systems Lab for their valuable discussions and insights.
    They also dedicate this article to the memory of Hind Rajab.
    This work has been supported by the AGAUR 2021 SGR 0075 grant and the Santa Fe Institute.
    J.\,P.\,M. was supported by grant PRE2020-091968, funded by MCIN/AEI (10.13039/501100011033), and co-funded by the ESF through the program ``Investing in your future''.
\end{acknowledgments}

\renewcommand{\bibsection}{\section*{References}}
\bibliography{bibliography}

@book{adamatzky2010physarum,
	title = {{Physarum machines: computers from slime mould}},
	author = {{Adamatzky, Andrew}},
	year = {2010},
	publisher = {{World Scientific}},
	volume = {74},
}

@book{sole2011phase,
  title={Phase transitions},
  author={Sol{\'e}, Ricard V},
  volume={16},
  year={2011},
  publisher={Princeton University Press}
}

@inproceedings{schumann2014bio,
  title={Bio-inspired game theory: The case of Physarum polycephalum},
  author={Schumann, Andrew and Pancerz, Krzysztof and Adamatzky, Andrew and Grube, Martin},
  booktitle={Proceedings of the 8th International Conference on Bioinspired Information and Communications Technologies},
  pages={9--16},
  year={2014}
}

@article{reid2016decision,
  title={Decision-making without a brain: how an amoeboid organism solves the two-armed bandit},
  author={Reid, Chris R and MacDonald, Hannelore and Mann, Richard P and Marshall, James AR and Latty, Tanya and Garnier, Simon},
  journal={Journal of The Royal Society Interface},
  volume={13},
  number={119},
  pages={20160030},
  year={2016},
  publisher={The Royal Society}
}

@article{dussutour2010amoeboid,
  title={Amoeboid organism solves complex nutritional challenges},
  author={Dussutour, Audrey and Latty, Tanya and Beekman, Madeleine and Simpson, Stephen J},
  journal={Proceedings of the national academy of sciences},
  volume={107},
  number={10},
  pages={4607--4611},
  year={2010},
  publisher={National Academy of Sciences}
}

@misc{levin2021uncovering,
  title={Uncovering cognitive similarities and differences, conservation and innovation},
  author={Levin, Michael and Keijzer, Fred and Lyon, Pamela and Arendt, Detlev},
  journal={Philosophical Transactions of the Royal Society B},
  volume={376},
  number={1821},
  pages={20200458},
  year={2021},
  publisher={The Royal Society}
}

@article{reid2016collective,
  title={Collective behaviour and swarm intelligence in slime moulds},
  author={Reid, Chris R and Latty, Tanya},
  journal={FEMS microbiology reviews},
  volume={40},
  number={6},
  pages={798--806},
  year={2016},
  publisher={Oxford University Press}
}

@article{deng2024development,
  title={The development of ecological systems along paths of least resistance},
  author={Deng, Jie and Cordero, Otto X and Fukami, Tadashi and Levin, Simon A and Pringle, Robert M and Sol{\'e}, Ricard and Saavedra, Serguei},
  journal={Current Biology},
  volume={34},
  number={20},
  pages={4813--4823},
  year={2024},
  publisher={Elsevier}
}

@article{kaila2008natural,
  title={Natural selection for least action},
  author={Kaila, Ville RI and Annila, Arto},
  journal={Proceedings of the Royal Society A: Mathematical, Physical and Engineering Sciences},
  volume={464},
  number={2099},
  pages={3055--3070},
  year={2008},
  publisher={The Royal Society London}
}

@article{lyon2006biogenic,
  title={The biogenic approach to cognition},
  author={Lyon, Pamela},
  journal={Cognitive Processing},
  volume={7},
  number={1},
  pages={11--29},
  year={2006},
  publisher={Springer}
}

@article{samuelson1974biological,
  title={A biological least-action principle for the ecological model of Volterra-Lotka},
  author={Samuelson, Paul A},
  journal={Proceedings of the National Academy of Sciences},
  volume={71},
  number={8},
  pages={3041--3044},
  year={1974}
}

@article{wang2022neural,
  title={Neural Network for Principle of Least Action},
  author={Wang, Beibei and Jackson, Shane and Nakano, Aiichiro and Nomura, Ken-ichi and Vashishta, Priya and Kalia, Rajiv and Stevens, Mark},
  journal={Journal of Chemical Information and Modeling},
  volume={62},
  number={14},
  pages={3346--3351},
  year={2022},
  publisher={ACS Publications}
}

@article{senn2024neuronal,
  title={A neuronal least-action principle for real-time learning in cortical circuits},
  author={Senn, Walter and Dold, Dominik and Kungl, Akos F and Ellenberger, Benjamin and Jordan, Jakob and Bengio, Yoshua and Sacramento, Jo{\~a}o and Petrovici, Mihai A},
  journal={ELife},
  volume={12},
  pages={RP89674},
  year={2024},
  publisher={eLife Sciences Publications Limited}
}

@incollection{grube2016physarum,
  title={Physarum, Quo Vadis?},
  author={Grube, Martin},
  booktitle={Advances in Physarum Machines: Sensing and Computing with Slime Mould},
  pages={23--35},
  year={2016},
  publisher={Springer}
}

@article{levin2023darwin,
  title={Darwin’s agential materials: evolutionary implications of multiscale competency in developmental biology},
  author={Levin, Michael},
  journal={Cellular and Molecular Life Sciences},
  volume={80},
  number={6},
  pages={142},
  year={2023},
  publisher={Springer}
}

@article{Lyon2015,
  author    = {Lyon, P.},
  title     = {The cognitive cell: bacterial behavior reconsidered},
  journal   = {Frontiers in Microbiology},
  year      = {2015},
  volume    = {6},
  pages     = {264},
  doi       = {10.3389/fmicb.2015.00264}
}

@article{BaluskaLevin2016,
  author    = {Balu{\v{s}}ka, F. and Levin, M.},
  title     = {On having no head: cognition throughout biological systems},
  journal   = {Frontiers in Psychology},
  year      = {2016},
  volume    = {7},
  pages     = {902},
  doi       = {10.3389/fpsyg.2016.00902}
}

@article{bajpai2025morphological,
  title={Morphological computational capacity of Physarum polycephalum},
  author={Bajpai, Suyash and Lucas-DeMott, Aviva and Murugan, Nirosha J and Levin, Michael and Kurian, Philip},
  journal={arXiv preprint arXiv:2510.19976},
  year={2025}
}

@article{awad2023survey,
  title={A survey on Physarum polycephalum intelligent foraging behaviour and bio-inspired applications},
  author={Awad, Abubakr and Pang, Wei and Lusseau, David and Coghill, George M},
  journal={Artificial Intelligence Review},
  volume={56},
  number={1},
  pages={1--26},
  year={2023},
  publisher={Springer}
}

@article{reid2023thoughts,
  title={Thoughts from the forest floor: a review of cognition in the slime mould Physarum polycephalum},
  author={Reid, Chris R},
  journal={Animal cognition},
  volume={26},
  number={6},
  pages={1783--1797},
  year={2023},
  publisher={Springer}
}

@article{baumgarten2010plasmodial,
  title={Plasmodial vein networks of the slime mold Physarum polycephalum form regular graphs},
  author={Baumgarten, Werner and Ueda, Tetsuo and Hauser, Marcus JB},
  journal={Physical Review E—Statistical, Nonlinear, and Soft Matter Physics},
  volume={82},
  number={4},
  pages={046113},
  year={2010},
  publisher={APS}
}

@article{takamatsu2001spatiotemporal,
  title={Spatiotemporal symmetry in rings of coupled biological oscillators of physarum plasmodial slime mold},
  author={Takamatsu, Atsuko and Tanaka, Reiko and Yamada, Hiroyasu and Nakagaki, Toshiyuki and Fujii, Teruo and Endo, Isao},
  journal={Physical Review Letters},
  volume={87},
  number={7},
  pages={078102},
  year={2001},
  publisher={APS}
}

@article{boisseau2016habituation,
  title={Habituation in non-neural organisms: evidence from slime moulds},
  author={Boisseau, Romain P and Vogel, David and Dussutour, Audrey},
  journal={Proceedings of the Royal Society B: Biological Sciences},
  volume={283},
  number={1829},
  pages={20160446},
  year={2016},
  publisher={The Royal Society}
}

@book{adamatzky2015atlas,
  title={Atlas of Physarum computing},
  author={Adamatzky, Andrew},
  year={2015},
  publisher={World Scientific}
}

@article{boussard2021adaptive,
  title={Adaptive behaviour and learning in slime moulds: the role of oscillations},
  author={Boussard, Aur{\`e}le and Fessel, Adrian and Oettmeier, Christina and Briard, L{\'e}a and D{\"o}bereiner, Hans-G{\"u}nther and Dussutour, Audrey},
  journal={Philosophical Transactions of the Royal Society B},
  volume={376},
  number={1820},
  pages={20190757},
  year={2021},
  publisher={The Royal Society}
}

@article{boussard2019memory,
  title={Memory inception and preservation in slime moulds: the quest for a common mechanism},
  author={Boussard, Aur{\`e}le and Delescluse, Julie and P{\'e}rez-Escudero, Alfonso and Dussutour, Audrey},
  journal={Philosophical Transactions of the Royal Society B},
  volume={374},
  number={1774},
  pages={20180368},
  year={2019},
  publisher={The Royal Society}
}

@article{liu2013physarum,
  title={Physarum optimization: A biology-inspired algorithm for the steiner tree problem in networks},
  author={Liu, Liang and Song, Yuning and Zhang, Haiyang and Ma, Huadong and Vasilakos, Athanasios V},
  journal={IEEE Transactions on Computers},
  volume={64},
  number={3},
  pages={818--831},
  year={2013},
  publisher={IEEE}
}

@article{adamatzky2012physarum,
  title={Physarum chip project: growing computers from slime mould.},
  author={Adamatzky, Andrew and Erokhin, Victor and Grube, Martin and Schubert, Theresa and Schumann, Andrew and others},
  journal={Int. J. Unconv. Comput.},
  volume={8},
  number={4},
  pages={319--323},
  year={2012}
}

@article{nakagaki2001path,
  title={Path finding by tube morphogenesis in an amoeboid organism},
  author={Nakagaki, Toshiyuki and Yamada, Hiroyasu and Toth, Agota},
  journal={Biophysical chemistry},
  volume={92},
  number={1-2},
  pages={47--52},
  year={2001},
  publisher={Elsevier}
}

@book{jones2015pattern,
  title={From pattern formation to material computation: multi-agent modelling of Physarum Polycephalum},
  author={Jones, Jeff},
  volume={15},
  year={2015},
  publisher={Springer}
}

@article{whiting2016towards,
  title={Towards a Physarum learning chip},
  author={Whiting, James GH and Jones, Jeff and Bull, Larry and Levin, Michael and Adamatzky, Andrew},
  journal={Scientific reports},
  volume={6},
  number={1},
  pages={19948},
  year={2016},
  publisher={Nature Publishing Group UK London}
}

@article{adamatzky2019brief,
  title={A brief history of liquid computers},
  author={Adamatzky, Andrew},
  journal={Philosophical Transactions of the Royal Society B},
  volume={374},
  number={1774},
  pages={20180372},
  year={2019},
  publisher={The Royal Society}
}

@article{lagzi2010maze,
  title={Maze solving by chemotactic droplets},
  author={Lagzi, Istv{\'a}n and Soh, Siowling and Wesson, Paul J and Browne, Kevin P and Grzybowski, Bartosz A},
  journal={Journal of the American Chemical Society},
  volume={132},
  number={4},
  pages={1198--1199},
  year={2010},
  publisher={ACS Publications}
}

@article{steinbock1995navigating,
  title={Navigating complex labyrinths: optimal paths from chemical waves},
  author={Steinbock, Oliver and T{\'o}th, {\'A}gota and Showalter, Kenneth},
  journal={Science},
  volume={267},
  number={5199},
  pages={868--871},
  year={1995},
  publisher={American Association for the Advancement of Science}
}

@article{dussutour2024flow,
  title={Flow-network adaptation and behavior in slime molds},
  author={Dussutour, Audrey and Arson, Chlo{\'e}},
  journal={Fungal Ecology},
  volume={68},
  pages={101325},
  year={2024},
  publisher={Elsevier}
}

@book{susskind2014theoretical,
  title={The theoretical minimum: what you need to know to start doing physics},
  author={Susskind, Leonard and Hrabovsky, George},
  year={2014},
  publisher={Basic Books}
}

@book{adamatzky2016advances,
	title = {{Advances in Physarum machines: Sensing and computing with slime mould}},
	author = {{Adamatzky, Andrew}},
	year = {2016},
	publisher = {{Springer}},
	volume = {21},
}

@article{valverde2004internet,
  title={Internet’s critical path horizon},
  author={Valverde, Sergi and Sol{\'e}, Ricard V},
  journal={The European Physical Journal B},
  volume={38},
  number={2},
  pages={245--252},
  year={2004},
  publisher={Springer}
}

@article{guimera2002optimal,
  title={Optimal network topologies for local search with congestion},
  author={Guimer{\`a}, Roger and D{\'\i}az-Guilera, Albert and Vega-Redondo, Fernando and Cabrales, Antonio and Arenas, Alex},
  journal={Physical review letters},
  volume={89},
  number={24},
  pages={248701},
  year={2002},
  publisher={APS}
}

@article{nakagaki2000maze,
  title={Maze-solving by an amoeboid organism},
  author={Nakagaki, Toshiyuki and Yamada, Hiroyasu and T{\'o}th, {\'A}gota},
  journal={Nature},
  volume={407},
  number={6803},
  pages={470--470},
  year={2000},
  publisher={Nature Publishing Group UK London}
}

@article{Bejan1997,
  author    = {Adrian Bejan},
  title     = {Constructal-theory network of conducting paths for cooling a heat generating volume},
  journal   = {Advances in Physics},
  volume    = {46},
  number    = {1},
  pages     = {1--21},
  year      = {1997},
  doi       = {10.1080/00018739700101488}
}

@article{tero2010rules,
  title={Rules for biologically inspired adaptive network design},
  author={Tero, Atsushi and Takagi, Seiji and Saigusa, Tetsu and Ito, Kentaro and Bebber, Dan P and Fricker, Mark D and Yumiki, Kenji and Kobayashi, Ryo and Nakagaki, Toshiyuki},
  journal={Science},
  volume={327},
  number={5964},
  pages={439--442},
  year={2010},
  publisher={American Association for the Advancement of Science}
}

@article{hasan2024filaments,
  title={Filaments of the slime mold cosmic web and how they affect galaxy evolution},
  author={Hasan, Farhanul and Burchett, Joseph N and Hellinger, Douglas and Elek, Oskar and Nagai, Daisuke and Faber, S\_M and Primack, Joel R and Koo, David C and Mandelker, Nir and Woo, Joanna},
  journal={The Astrophysical Journal},
  volume={970},
  number={2},
  pages={177},
  year={2024},
  publisher={IOP Publishing}
}

@article{reid2013solving,
  title={Solving the Towers of Hanoi--how an amoeboid organism efficiently constructs transport networks},
  author={Reid, Chris R and Beekman, Madeleine},
  journal={Journal of Experimental Biology},
  volume={216},
  number={9},
  pages={1546--1551},
  year={2013},
  publisher={Company of Biologists}
}

@article{katifori2010damage,
  title={Damage and fluctuations induce loops in optimal transport networks},
  author={Katifori, Eleni and Sz{\"o}ll{\H{o}}si, Gergely J and Magnasco, Marcelo O},
  journal={Physical review letters},
  volume={104},
  number={4},
  pages={048704},
  year={2010},
  publisher={APS}
}

@article{hu2013adaptation,
  title={Adaptation and optimization of biological transport networks},
  author={Hu, Dan and Cai, David},
  journal={Physical review letters},
  volume={111},
  number={13},
  pages={138701},
  year={2013},
  publisher={APS}
}

@article{bohn2007structure,
  title={Structure, scaling, and phase transition in the optimal transport network},
  author={Bohn, Steffen and Magnasco, Marcelo O},
  journal={Physical review letters},
  volume={98},
  number={8},
  pages={088702},
  year={2007},
  publisher={APS}
}

@article{forrest1990emergent,
  title={Emergent computation: self-organizing, collective, and cooperative phenomena in natural and artificial computing networks: introduction to the proceedings of the ninth annual CNLS conference},
  author={Forrest, Stephanie},
  journal={Physica D: Nonlinear Phenomena},
  volume={42},
  number={1-3},
  pages={1--11},
  year={1990},
  publisher={Elsevier}
}

@article{le2024physarum,
  title={Physarum polycephalum: Smart network adaptation},
  author={Le Verge-Serandour, Mathieu and Alim, Karen},
  journal={Annual Review of Condensed Matter Physics},
  volume={15},
  year={2024},
  publisher={Annual Reviews}
}

@book{barabasi2016network,
  title        = {Network Science},
  author       = {Barab{\'a}si, Albert-L{\'a}szl{\'o}},
  year         = {2016},
  publisher    = {Cambridge University Press},
  address      = {Cambridge, UK},
  url          = {http://networksciencebook.com},
  isbn         = {9781107076266}
}

@article{barthelemy2011spatial,
  title={Spatial networks},
  author={Barth{\'e}lemy, Marc},
  journal={Physics reports},
  volume={499},
  number={1-3},
  pages={1--101},
  year={2011},
  publisher={Elsevier}
}

@article{kelly1991network,
  title={Network routing},
  author={Kelly, Frank P},
  journal={Philosophical Transactions of the Royal Society of London. Series A: Physical and Engineering Sciences},
  volume={337},
  number={1647},
  pages={343--367},
  year={1991},
  publisher={The Royal Society London}
}

@article{i2001topology,
  title={Topology of technology graphs: Small world patterns in electronic circuits},
  author={i Cancho, Ramon Ferrer and Janssen, Christiaan and Sol{\'e}, Ricard V},
  journal={Physical Review E},
  volume={64},
  number={4},
  pages={046119},
  year={2001},
  publisher={APS}
}

@article{pinero2019statistical,
  title={Statistical physics of liquid brains},
  author={Pi{\~n}ero, Jordi and Sol{\'e}, Ricard},
  journal={Philosophical Transactions of the Royal Society B},
  volume={374},
  number={1774},
  pages={20180376},
  year={2019},
  publisher={The Royal Society}
}

@book{rojas2013neural,
  title={Neural networks: a systematic introduction},
  author={Rojas, Ra{\'u}l},
  year={2013},
  publisher={Springer Science \& Business Media}
}

@article{farmer1990rosetta,
  title={A Rosetta stone for connectionism},
  author={Farmer, J Doyne},
  journal={Physica D: Nonlinear Phenomena},
  volume={42},
  number={1-3},
  pages={153--187},
  year={1990},
  publisher={Elsevier}
}

@article{fernandez2025foraging,
  title={Foraging ants as liquid brains: Movement heterogeneity shapes collective efficiency},
  author={Fern{\'a}ndez-L{\'o}pez, Pol and Oro, Daniel and Lloret-Cabot, Roger and Genovart, Meritxell and Garriga, Joan and Bartumeus, Frederic},
  journal={Proceedings of the National Academy of Sciences},
  volume={122},
  number={31},
  pages={e2506930122},
  year={2025},
  publisher={National Academy of Sciences}
}

@book{hertz2018introduction,
  title={Introduction to the theory of neural computation},
  author={Hertz, J and Krogh, A and Palmer, R},
  year={2018},
  publisher={Crc Press}
}

@article{hopfield1982neural,
  title={Neural networks and physical systems with emergent collective computational abilities.},
  author={Hopfield, John J},
  journal={Proceedings of the national academy of sciences},
  volume={79},
  number={8},
  pages={2554--2558},
  year={1982}
}

@article{deneubourg1983probabilistic,
	title = {{Probabilistic behaviour in ants: a strategy of errors?}},
	author = {{Deneubourg, Jean-Louis and Pasteels, Jacques M and Verhaeghe, Jean-Claude}},
	year = {1983},
	journal = {{Journal of theoretical Biology}},
	publisher = {{Elsevier}},
	volume = {105},
	number = {2},
	pages = {259--271},
}

@article{deneubourg1990self,
	title = {{The self-organizing exploratory pattern of the argentine ant}},
	author = {{Deneubourg, J-L and Aron, Serge and Goss, Simon and Pasteels, Jacques M}},
	year = {1990},
	journal = {{Journal of insect behavior}},
	publisher = {{Springer}},
	volume = {3},
	number = {2},
	pages = {159--168},
}

@article{dussutour2021learning,
	title = {{Learning in single cell organisms}},
	author = {{Dussutour, Audrey}},
	year = {2021},
	journal = {{Biochemical and Biophysical Research Communications}},
	publisher = {{Elsevier}},
	volume = {564},
	pages = {92--102},
}

@article{oborny2019plant,
	title = {{The plant body as a network of semi-autonomous agents: a review}},
	author = {{Oborny, Beata}},
	year = {2019},
	journal = {{Philosophical Transactions of the Royal Society B}},
	publisher = {{The Royal Society}},
	volume = {374},
	number = {1774},
	pages = {20180371},
}

@article{Olle2016Morphospace,
	title = {{A morphospace for synthetic organs and organoids: the possible and the actual}},
	author = {{Oll\'{e}-Vila,  Aina and Duran-Nebreda,  Salva and Conde-Pueyo,  N\'{u}ria and Monta\~{n}ez,  Ra\'{u}l and Sol\'{e},  Ricard}},
	year = {2016},
	month = apr,
	journal = {{Integrative Biology}},
	publisher = {{Oxford University Press (OUP)}},
	volume = {8},
	number = {4},
	pages = {485–503},
	doi = {10.1039/c5ib00324e},
	issn = {1757-9708},
	url = {http://dx.doi.org/10.1039/C5IB00324E},
}

@article{sole2000pattern,
	title = {{Pattern formation and optimization in army ant raids}},
	author = {{Sol{\'e}, Ricard V and Bonabeau, Eric and Delgado, Jordi and Fern{\'a}ndez, Pau and Mar{\'\i}n, Jesus}},
	year = {2000},
	journal = {{Artificial life}},
	publisher = {{MIT Press}},
	volume = {6},
	number = {3},
	pages = {219--226},
}

@article{sole2019liquid,
	title = {{Liquid brains, solid brains}},
	author = {{Sol{\'e}, Ricard and Moses, Melanie and Forrest, Stephanie}},
	year = {2019},
	month = apr,
	journal = {{Philosophical Transactions of the Royal Society B}},
	publisher = {{The Royal Society}},
	volume = {374},
	number = {1774},
	pages = {20190040},
	issn = {0962-8436},
}

@article{sole2024open,
	title = {{Open problems in synthetic multicellularity}},
	author = {{Sol\'{e},  Ricard and Conde–Pueyo,  N\'{u}ria and Pla–Mauri,  Jordi and Garcia–Ojalvo,  Jordi and Montserrat,  Nuria and Levin,  Michael}},
	year = {2024},
	month = dec,
	journal = {{npj Systems Biology and Applications}},
	publisher = {{Springer Science and Business Media LLC}},
	volume = {10},
	number = {1},
	doi = {10.1038/s41540-024-00477-8},
	issn = {2056-7189},
	url = {http://dx.doi.org/10.1038/s41540-024-00477-8},
}

@article{stern2023learning,
	title = {{Learning without neurons in physical systems}},
	author = {{Stern, Menachem and Murugan, Arvind}},
	year = {2023},
	month = mar,
	journal = {{Annual Review of Condensed Matter Physics}},
	publisher = {{Annual Reviews}},
	volume = {14},
	number = {1},
	pages = {417--441},
	doi = {10.1146/annurev-conmatphys-040821-113439},
	issn = {1947-5454},
	url = {https://www.annualreviews.org/doi/pdf/10.1146/annurev-conmatphys-040821-113439},
}

@article{Tero2007,
	title = {{A mathematical model for adaptive transport network in path finding by true slime mold}},
	author = {{Tero,  Atsushi and Kobayashi,  Ryo and Nakagaki,  Toshiyuki}},
	year = {2007},
	month = feb,
	journal = {{Journal of Theoretical Biology}},
	publisher = {{Elsevier BV}},
	volume = {244},
	number = {4},
	pages = {553–564},
	doi = {10.1016/j.jtbi.2006.07.015},
	issn = {0022-5193},
	url = {http://dx.doi.org/10.1016/j.jtbi.2006.07.015},
}

@article{trewavas2003aspects,
	title = {{Aspects of plant intelligence}},
	author = {{Trewavas, Anthony}},
	year = {2003},
	journal = {{Annals of botany}},
	publisher = {{Oxford University Press}},
	volume = {92},
	number = {1},
	pages = {1--20},
}

\end{document}